\documentclass[prd,showpacs,preprintnumbers,nofootinbib]{revtex4} % Physical Review D

\usepackage{graphicx}% Include figure files
\usepackage{dcolumn}% Align table columns on decimal point
\usepackage{bm}% bold math
\usepackage{latexsym}
\usepackage{amsfonts}
\usepackage{amssymb}
\usepackage{amsmath}
\begin{document}

%\preprint{}

\title{
Renormalized Newtonian Cosmic Evolution 
 with Primordial  Non-Gaussianity 
}

\author{Keisuke Izumi}
\email{ksuke@tap.scphys.kyoto-u.ac.jp}
\author{Jiro Soda}
\email{jiro@tap.scphys.kyoto-u.ac.jp}
\affiliation{
 Department of Physics,  Kyoto University, Kyoto 606-8501, Japan
}%

\date{\today}% It is always \today, today,

%===============================================================%
%************************* ABSTRACT ****************************%
%===============================================================%
\begin{abstract}
 We study  Newtonian cosmological perturbation theory
 from a field theoretical point of view.
 We derive a path integral representation for the cosmological
 evolution of stochastic fluctuations. 
 Our main result is the closed form of the generating functional
 valid for any initial statistics. Moreover, 
 we extend the renormalization group method proposed by Mataresse and Pietroni
 to the case of primordial non-Gaussian density and velocity fluctuations.  
 As an application, we calculate the nonlinear propagator
 and examine how the non-Gaussianity affects the memory of cosmic fields
 to their initial conditions. It turns out that the non-Gaussianity
 affect the nonlinear propagator. 
 In the case of positive skewness, the onset of the nonlinearity is
 advanced with a given comoving wavenumber. 
 On the other hand, the negative skewness gives the opposite result.
\end{abstract}

\maketitle
%===============================================================%
%************************ SECTION I **************************%
%===============================================================%
\section{Introduction}
 
 The large scale structure in the universe has evolved from
 primordial fluctuations according to the gravitational
 instability. In the standard scenario of the structure formation,
the primordial fluctuations are created quantum mechanically
during the inflationary stage in the early universe. 
After exiting the horizon, the fluctuations are evolved linearly; which
is well described by relativistic linear perturbation theory.
Eventually the fluctuations re-enter into the  horizon.
After that, it is sufficient to treat the evolution of fluctuations
by means  of Newtonian gravity. 
Due to the Jeans instability, at some point, the density fluctuations
 become nonlinear. In this stage, usually we resort to
the N-body simulations. However, since the numerical simulations
are time consuming, the analytical calculation
of the nonlinear evolution is still desired. 
The standard perturbative expansion method is developed for this purpose.
In the quasi-nonlinear regime, the perturbative approach 
was successful~\cite{Suto:1990wf,Makino:1991rp,Jain:1993jh,
Scoccimarro:1995if,Sahni:1995rm,Matsubara:2000mw,Bernardeau:2001qr}.
To obtain more accurate results, however, the non-perturbative analytic method 
would be necessary.

Recently, Crocce and Scoccimarro have developed a new formalism to study
the large scale structure~\cite{Crocce:2005xy}.
 They described the perturbative solution
by Feynman diagrams and identified three fundamental objects:
the initial conditions, the vertex, and the propagator.
They have found that the renormalization of the propagator is the
most important one. Based on this finding,
they have observed that, due to the rapid fall off of the nonlinear propagator,
the memory of the cosmic fields to their initial conditions will be
lost soon in the nonlinear regime. 

Following their work, Matarrese and Pietroni reformulated the 
cosmological perturbation theory from the path integral point of view
and developed the renormalization group (RG) techniques 
in cosmology~\cite{Matarrese:2007aj,Matarrese:2007wc}
 ( see \cite{McDonald:2006hf} for a slightly different approach). 
Matarrese and Pietroni have applied their formalism to the baryon acoustic oscillations
(BAO) which takes place around the 
scale $k\sim 0.1 {\rm Mpc}^{-1}$~\cite{Eisenstein:2005su}.
On these scales, the nonlinear effects are relevant~\cite{Huff:2006gs,Jeong:2006xd}.
They have found that the renormalization group method is useful
to predict the BAO feature. Crocce and Scoccimarro have used their
graphical approach to discuss BAO and found the renormalized
perturbation approach gives a good agreements with results of
numerical simulations~\cite{Crocce:2007dt}.  This result
is further confirmed by Nishimichi et al.~\cite{Nishimichi:2007xt}.

These authors have discussed only the Gaussian initial conditions.
Recently, however, it has been realized that 
the primordial non-Gaussianity can be produced in the inflationary scenario.
If so, it is important to  give a renormalization group formalism
for the non-Gaussian initial conditions. Conventionary, the non-Gaussian
curvature perturbation $\Phi$ is characterized by the following form
\begin{eqnarray}
\Phi ({\bf x}) = \Phi_g ({\bf x}) 
+ f_{NL} (\Phi_g^2 ({\bf x}) - \langle \Phi_g^2 \rangle ) ,
\end{eqnarray}
where $\Phi_g$ is the Gaussian field and 
$f_{NL}$ is the parameter to represent a deviation from the Gaussianity.
There are several possible observational tests to constrain 
the non-Gaussianity~\cite{Koyama:1999fc,Willick:1999ty,Robinson:1999wh,
Matarrese:2000iz,Scoccimarro:2000qg,Seto:2001mg,
Scoccimarro:2003wn,Hikage:2006fe,Sefusatti:2006eu}.
The most stringent limit comes from WMAP and the result is 
$-58<f_{NL}<134$~\cite{Komatsu:2003fd}.
Planck and other tests will give a more stringent limit or 
detect the non-Gaussianity~\cite{Bartolo:2004if}.

The purpose of this paper is therefore to extend the analysis by Matarrese and Pietroni
to the non-Gaussian initial conditions. 
Starting from the generating functional of the multi-point functions,
we derive the path integral representation of the cosmic evolution of
the cosmic fields.  In contrast to the previous work, 
the non-Gaussianity is incorporated into the field theoretical scheme. 
In particular, we obtain the formula for the generating functional which
allows us to use the Feynman diagram method to calculate
various statistical quantities characterizing the large scale structure in
the universe. We also derive the RG equation for the effective action.
As an application, we calculate the nonlinear propagator and examine if the
memory of the cosmic fields to their initial conditions 
has the tendency to be kept by the non-Gaussianity or not.

The organization of this paper is as follows.
In section II, we review the basic equations of motion describing
the evolution of the cosmic fields.
In section III, we develop the general field theoretical framework
so that the non-Gaussianity can be incorporated into the scheme.
We have successfully calculated the generating functional which gives
Feynman rules. 
We also develop the renormalization group method in this context. 
In section IV, we apply our formalism to calculate the
nonlinear propagator for the non-Gaussian initial conditions
 and examine the effect of non-Gaussianity on the nonlinear scales. 
The final section is devoted to  the conclusion.

%===============================================================%
%************************ SECTION II **************************%
%===============================================================%
\section{Basic equations for cosmic fields}

In this section, we review the standard Newtonian
cosmological perturbation theory. 
Here, we consider the Einstein-de Sitter universe for simplicity. 
Of course, it is possible to extend our analysis to other cosmological 
models. 

First of all, let us consider the homogeneous cosmological background
spacetime. 
Taking the conformal time and assuming the flat space, 
we can write down the metric
\begin{eqnarray}
   ds^2 = a^2(\tau) \left[ -d\tau^2 + \delta_{ij} dx^i dx^j \right] \ .
\end{eqnarray}
The cosmological scale factor $a$ is determined by solving
FRW equations
\begin{eqnarray}
  {\cal H}^2 = \frac{8\pi G}{3}a^2 \rho_0 \ , \quad
  {\cal H}' = - \frac{4\pi G}{3} a^2 \rho_0   \ ,
\end{eqnarray}
where $\rho_0$ is the averaged density field
and we have defined ${\cal H}= da/d\tau/a = a' /a$. 

Now, let us consider the inhomogeneous distribution of the matter. 
The evolution of the total matter density is determined by the
gravity including the effect of cosmic expansion.
The actual density $\rho({\bf x} ,\tau )$ 
is deviated from the averaged density $\rho_0 (\tau) $
Let us define the density fluctuation as
\begin{eqnarray}
\delta ({\bf x} ,\tau ) 
= \frac{\rho({\bf x} ,\tau ) -\rho_0 (\tau)}{\rho_0 (\tau)}
= \int d^3 k \delta ({\bf k},\tau ) e^{-i{\bf k}\cdot{\bf x}} \ .
\end{eqnarray}
It obeys the equation of continuity and the peculiar velocity ${\bf v}$
is determined by the Euler equation in the presence of gravitational potential $\phi$.
The gravitational potential itself is governed by the Poisson equation.
Thus, equations of motion for the cosmic fields, $\delta$, ${\bf v}$, and $\phi$,
 are given by
\begin{eqnarray}
  && \frac{\partial \delta}{\partial \tau} 
  + \nabla \cdot \left[ \left(1+\delta \right) {\bf v} \right]=0,  \\
 && \frac{\partial {\bf v}}{\partial \tau} + {\cal H} {\bf v}
 + \left( {\bf v}\cdot \nabla \right) {\bf v} = - \nabla \phi, \\
 && \nabla^2 \phi = \frac{3}{2}{\cal H}^2 \delta \ .
\end{eqnarray}
On large scales, the assumption that the peculiar velocity is irrotational
would be valid. Then, defining $\theta = \nabla \cdot {\bf v}$, 
we obtain the relation ${\bf v}({\bf k},\tau) =i{\bf k} \theta ({\bf k},\tau) /k^2$.
After eliminating the gravitational potential, we obtain
equations of motion in the Fourier space
\begin{eqnarray}
&& \frac{\partial \delta ({\bf k} ,\tau) }{\partial \tau} 
+ \theta ({\bf k},\tau)
+ \int d^3 {\bf q} d^3{\bf p} \delta_D ({\bf k} - {\bf q} -{\bf p})
  \alpha ({\bf q},{\bf p}) \theta ({\bf q}, \tau ) \delta ({\bf p} ,\tau ) =0 , \\
 && \frac{\partial \theta ({\bf k} ,\tau) }{\partial \tau}
  + {\cal H} \theta ({\bf k},\tau) + \frac{3}{2}{\cal H}^2 \delta ({\bf k},\tau)
  \nonumber\\
&&  \qquad + \int d^3 {\bf q} d^3{\bf p} \delta_D ({\bf k} - {\bf q} -{\bf p})
  \beta ({\bf q},{\bf p}) \theta ({\bf q}, \tau ) \theta ({\bf p} ,\tau ) =0 ,
\end{eqnarray}
where we have defined the Dirac delta function $\delta_D $ and 
\begin{eqnarray}
  \alpha ( {\bf q},{\bf p}) = \frac{({\bf p} +{\bf q})\cdot {\bf q}}{q^2} \ , \quad
  \beta ( {\bf q},{\bf p}) 
          = \frac{({\bf p} +{\bf q})^2 {\bf p}\cdot {\bf q}}{2 p^2 q^2}
\end{eqnarray}
characterize the nonlinear gravitational coupling. 

It is now easy to solve the above equations perturbatively.
Given the expansion  
\begin{eqnarray}
  \delta ({\bf k},\tau) = \sum_{n=1}^\infty  \delta^{(n)} ({\bf k},\tau) \ , \quad
   \theta ({\bf k},\tau) = \sum_{n=1}^\infty  \theta^{(n)} ({\bf k},\tau) \ ,
\end{eqnarray}
the iterative solutions can be explicitly written down as functional of the
initial fields~\cite{Bernardeau:2001qr}. 
Once the probability functional for the initial fields are given,
one can calculate expectation values of product of these fields. In fact, 
this standard formalism has been utilized to calculate the quasi-nonlinear
evolution of the power spectrum, bispectrum, and other statistics such as 
genus statistics~\cite{Sahni:1995rm,Matsubara:2000mw,Bernardeau:2001qr}.
 In the nonlinear regime, however, further analytical
tools are required to give more accurate results. 
This is crucial for the prediction of BAO feature, for example. 
In the next section, we introduce a useful approach to this end.  
  
%===============================================================%
%************************ SECTION III **************************%
%===============================================================%
\section{Path Integral Formalism and Renormalization Group}

Now, we proceed to formulate the cosmological perturbation
theory in the field theoretical manner.
Doing so, we can utilize the idea invented in the field theory.
In particular, the renormalization group method turns out to be useful.

As is suggested by Crocce and Scoccimarro~\cite{Crocce:2005xy}, 
the equations of motion can be rewritten in a convenient form
by defining a two-component vector
\begin{eqnarray}
\varphi_a ({\bf k},\tau) \equiv   \left( 
   \begin{array}{c}
   \varphi_1 ({\bf k},\tau) \\
   \varphi_2 ({\bf k},\tau)
   \end{array}
   \right)
   \equiv e^{-\eta}
    \left( 
   \begin{array}{c}
   \delta ({\bf k},\tau) \\
   - \theta ({\bf k},\tau) /{\cal H} 
   \end{array}
   \right) ,
\end{eqnarray}
where the index $a = 1,2$ and $\eta = \log a (\tau) /a_{\rm in}$
denotes the e-folding number. The initial scale factor $a_{\rm in}$
can be taken arbitrarily. 
We also define vertex  $\gamma_{abc}$ with
\begin{eqnarray}
  \gamma_{121} ({\bf k} , {\bf p} , {\bf q} )
  = \frac{1}{2} \delta_D ({\bf k} + {\bf p} + {\bf q} ) \alpha ({\bf p},{\bf q}) 
  =\gamma_{112} ({\bf k} , {\bf q} , {\bf p} )
  \ , \quad
  \gamma_{222} ({\bf k} , {\bf p} , {\bf q} )
  =  \delta_D ({\bf k} + {\bf p} + {\bf q} ) \beta ({\bf p},{\bf q})
\end{eqnarray}
and other components are zero. 
Using the above definitions, we obtain the equation
\begin{eqnarray}
  \left( \delta_{ab} \partial_\eta + \Omega_{ab} \right) \varphi_b ({\bf k} ,\eta)
  = e^{\eta} \gamma_{abc} ({\bf k} , - {\bf p}, -{\bf q} )
  \varphi_b ({\bf p} ,\eta) \varphi_c ({\bf q} ,\eta) , \label{EOM}
\end{eqnarray}
where $\Omega_{ab}$ are components of the matrix
\begin{eqnarray}
\Omega = \left( 
   \begin{array}{cc}
   1 & -1\\
   -3/2 & 3/2
   \end{array}
   \right) .
\end{eqnarray}
Here, we have used the Einstein's sum rule
\begin{eqnarray}
   \varphi_a (-{\bf k}, 0) \varphi_b ({\bf k}, 0)
   \equiv  \int d^3 k \varphi_a (-{\bf k}, 0) \varphi_b ({\bf k}, 0) \ .
\end{eqnarray}
From now on, we should understand this convention is used
when the same wavenumber vector appear twice in the same term. 

First, we consider the linear theory. 
The growing and the  decaying modes can be written as
$u_a$ and $v_a \exp(-5\eta /2 )$, respectively. Here, we have defined
two basis vectors
\begin{eqnarray}
u_a = \left(
  \begin{array}{c}
    1   \\
     1  \\
  \end{array}
\right) 
\qquad
\mbox{and}
\qquad
v_a = \left(
  \begin{array}{c}
    1   \\
   -3/2  \\
  \end{array}
\right) .
\end{eqnarray}
Let us define the linear propagator as
\begin{eqnarray}
   \left( \delta_{ab} \partial_\eta + \Omega_{ab} \right) g_{bc} (\eta_a -\eta_b ) 
   = \delta_{ac} \delta_D (\eta_a -\eta_b ) \ .
\end{eqnarray}
Given the growing and decaying mode functions, 
we can write down the causal propagator $g_{ab}$ as
\begin{eqnarray}
g_{ab} (\eta_a , \eta_b )
= \left\{ u_a \hat u_b + v_a \hat v_b \exp{(-5/2(\eta_a -\eta_b))} \right\}
\theta(\eta_a -\eta_b),
\end{eqnarray}
where $\hat u_a$ and $\hat v_a$ are dual vectors of $u_a$ and $v_a$, 
\begin{eqnarray}
\hat u_a = \frac{1}{5} (3,2) 
\qquad
\mbox{and}
\qquad
\hat v_a = \frac{2}{5}(1,-1).
\end{eqnarray}
Note that we have the matrix components
\begin{eqnarray}
u_a \hat u_b = \frac{1}{5} \left( 
   \begin{array}{cc}
   3 & 2\\
   3 & 2
   \end{array}
   \right) \ , \quad 
v_a \hat v_b = \frac{1}{5} \left( 
   \begin{array}{cc}
   2 & -2\\
   -3 & 3
   \end{array}
   \right)  .
\end{eqnarray}
Using this linear propagator, we can formally solve the equation (\ref{EOM}) as
\begin{eqnarray}
   \varphi_a ({\bf k}, \eta_a ) = g_{ab} (\eta_a , 0 ) \varphi_b ({\bf k}, 0)
   + \int d\eta_b g_{ab} (\eta_a , \eta_b )
   e^{\eta} \gamma_{bcd} ({\bf k} , - {\bf p}, -{\bf q} )
  \varphi_c ({\bf p} ,\eta_b ) \varphi_d ({\bf q} ,\eta_b ) \ .
\end{eqnarray}
Iteration gives the perturbative solutions.
There are three building blocks, the initial field $\varphi ({\bf k},0)$,
 the linear propagator $g_{ab}$, and the vertex $\gamma_{abc}$. 
 The initial field is usually assumed to have Gaussian statistics
 characterized by the linear power spectrum
\begin{eqnarray}
   < \varphi ({\bf k},0) \varphi ({\bf k}',0)> 
   = (2\pi)^3 \delta_D ({\bf k}+{\bf k}') P_{ab}^L \ .
\end{eqnarray}
The nonlinear power spectrum can be calculated using the graphical
method as was first demonstrated by Crocce and Scoccimarro~\cite{Crocce:2005xy}.
In particular, the renormalization method is utilized to
perform the partial sum of diagrams. It turned out the 
approximation motivated by the renormalization method
was quite successful~\cite{Afshordi:2006ch}. 
In the next subsection, we consider more general statistics for
the initial field $\varphi_a ({\bf k} , 0)$.

\subsection{Path Integral Representation}

What we are interested in are the statistical quantities characterizing
the large scale structure in the universe. 
The statistics of primordial cosmic fields $\varphi_a ({\bf k},0)$ are determined
by the initial probability functional $P[\varphi_a ({\bf k}, 0) ]$.
To calculate desired quantities, we need to solve the nonlinear evolution
equations and obtain the solution as a function of the initial fields. 
Namely, we have the statistics and the dynamics  to be considered. 
More precisely, we want to calculate
\begin{eqnarray}
  && \left< \exp i \int d\eta J_a (-{\bf k}, \eta) 
  \varphi_a ({\bf k},\eta ; \varphi ({\bf k},0)) \right> \nonumber\\
&&  \qquad = \int d \varphi_a ({\bf k}, 0) P[\varphi_a ({\bf k}, 0) ] 
   \exp i \int d\eta J_a (-{\bf k}, \eta) \varphi_a ({\bf k}, \eta ; \varphi_a ({\bf k}, 0)) ,
   \label{key}
\end{eqnarray}
where $P[\varphi_a ({\bf k}, 0) ]$ is the general probability functional 
for the initial field $\varphi_a ({\bf k}, 0)$ and 
 $\varphi_a ({\bf k}, \eta ; \varphi_a ({\bf k}, 0))$ is the solution of Eq.(\ref{EOM})
with the initial condition $\varphi_a ({\bf k}, 0)$. This is a generating functional
for multi-point correlation functions.
Here, we shall combine  the statistics and the dynamics in a unified framework. 
This can be achieved by the field theoretical path integral method. 
It is the path integral representation of the problem
 which can be used to perform the non-perturbative approximation.

To derive the path integral representation for the cosmic fields
starting from the expression (\ref{key}), 
we introduce an auxiliary field $\varphi_a ({\bf k}, \eta)$ as 
\begin{eqnarray}
 \int d \varphi_a ({\bf k},0) \int D\varphi_a ({\bf k}, \eta) P[\varphi_a ({\bf k}, 0) ] 
   \delta_D ( \varphi_a ({\bf k}, \eta) - \varphi_a ({\bf k}, \eta ; \varphi ({\bf k}, 0)) )
   \exp i \int d\eta J_a (-{\bf k}, \eta) \varphi_a ({\bf k}, \eta )   \ ,
\end{eqnarray}
where we used the Dirac delta function $\delta_D$. 
To separate the dynamics from the statistics, we use the operator
${\bf L}$  defined by
\begin{eqnarray}
{\bf L} \varphi_a 
=  \left( \delta_{ab} \partial_\eta + \Omega_{ab} \right) \varphi_b ({\bf k} ,\eta)
  - e^{\eta} \gamma_{abc} ({\bf k} , - {\bf p}, -{\bf q} )
  \varphi_b ({\bf p} ,\eta) \varphi_c ({\bf q} ,\eta) \ .
\end{eqnarray}  
Then, we get 
\begin{eqnarray}
  \int d \varphi_a ({\bf k}, 0) \int D\varphi_a ({\bf k}, \eta) P[\varphi_a ({\bf k}, 0) ] 
   \delta_D ( {\bf L} \varphi_a ({\bf k}, \eta) - \varphi_a ({\bf k}, 0) \delta_D (\eta) )
   \exp i \int d\eta J_a (-{\bf k}, \eta) \varphi_a ({\bf k}, \eta ) ,
\end{eqnarray}
where  we have used the fact $\det {\bf L} =1$ valid
for the causal boundary conditions~\cite{Valageas:2006bi}. Here, it is convenient to
introduce another auxiliary field $\chi_a $ to exponentiate the Dirac delta function.
This procedure is crucial to integrate out the initial field $\varphi_a ({\bf k}, 0)$.
The result becomes
\begin{eqnarray}   
&&  \int d \varphi_a ({\bf k}, 0) \int D\varphi_a ({\bf k}, \eta) 
     D\chi_a  ({\bf k}, \eta) P[\varphi_a ({\bf k}, 0) ]    \nonumber\\
 && \quad \times  \exp \left[ i\int d\eta \chi_a (-{\bf k}, \eta) L 
   \varphi_a ({\bf k}, \eta) -i \chi_a (-{\bf k}, 0) \varphi_a ({\bf k}, 0)
   + i \int d\eta J_a (-{\bf k}, \eta) \varphi_a ({\bf k}, \eta ) \right] \nonumber\\
&=&  \int D\varphi_a ({\bf k}, \eta) D\chi_a  ({\bf k}, \eta) 
          e^{C\left[\chi_a ({\bf k}, 0) \right] }
   \exp \left[ i\int d\eta \chi_a (-{\bf k}, \eta) {\bf L} \varphi_a ({\bf k}, \eta) 
   + i \int d\eta J_a (-{\bf k}, \eta) \varphi_a ({\bf k}, \eta ) \right]  ,             
\end{eqnarray}
where we have defined the cumulant functional $C[\chi_a ]$ by
\begin{eqnarray}
    e^{C\left[\chi_a ({\bf k}, 0) \right] } 
    = \int d \varphi_a ({\bf k}, 0)  P[\varphi_a ({\bf k}, 0) ]
     e^{-i \chi_a (-{\bf k}, 0) \varphi_a ({\bf k}, 0)} .
\end{eqnarray}
The field $\chi_a ({\bf k}, 0)$ is defined as the boundary value of the
field $\chi_a ({\bf k}, \eta )$.
The cumulant functional completely characterize the initial statistics.
It is possible to expand the cumulant as
\begin{eqnarray}
C[\chi_a ({\bf k}, 0) ] 
= (2\pi)^3\left(- \frac{1}{2} \chi_a (-{\bf k}, 0) P_{ab} \chi_b ({\bf k}, 0) 
  +\frac{i}{6} B_{abc}\left({\bf k}_1  ,{\bf k}_2 , {\bf k}_3  \right)
   \chi_a (-{\bf k}_1 , 0) \chi_b (-{\bf k}_2 , 0) \chi_c (-{\bf k}_3 , 0) + \cdots \right) \ .
\end{eqnarray}
The first term is the power spectrum for the initial field. 
In the Gaussian case, only this term exists.
In general, subsequent terms follow. The second term is called the bispectrum

In the above action, the auxiliary field $\chi_a$ is introduced as a field. 
Hence, it is natural to add the source for this field. 
Thus, we have the final path integral expression for the cosmological
fluctuations with the general statistics:
\begin{eqnarray}
Z[J_a ,K_b ] &=& \int D\varphi_a ({\bf k}, \eta) D\chi_a  ({\bf k}, \eta) 
                e^{C[\chi_a ({\bf k}, 0) ] }
   \exp \left[ i\int d\eta \chi_a (-{\bf k}, \eta) {\bf L} 
   \varphi_a ({\bf k}, \eta) \right. \nonumber\\
  && \left. \qquad + i \int d\eta J_a (-{\bf k}, \eta) \varphi_a ({\bf k}, \eta ) 
  + i \int d\eta K_a (-{\bf k}, \eta) \chi_a ({\bf k}, \eta ) \right] .
\end{eqnarray}
From the field theoretical point of view, $C[\chi_a ({\bf k}, 0)]$
can be regarded as the boundary action on the initial hypersurface $\eta=0$. 
In this sense, the field $\chi_a ({\bf k}, 0)$ 
is associated with the initial conditions.

\subsection{Generating functional}

Thanks to the auxiliary field $\chi_a$, we can perform the path integral
completly. 
First of all, let us extract the nonlinear interaction part 
\begin{eqnarray}
Z[J_a ,K_b ] &=& \exp i S_{int} [-i \frac{\delta}{ \delta J_a} , -i \frac{\delta}{\delta K_b}]
       \int D\varphi_a ({\bf k}, \eta) D\chi_a  ({\bf k}, \eta) 
       e^{C[\chi_a ({\bf k}, 0) ] }   \nonumber \\
    &&    \times \exp \left[ i\int d\eta \chi_a (-{\bf k}, \eta) 
    {\bf L}_{lin} \varphi_a ({\bf k}, \eta) 
   + i \int d\eta J_a (-{\bf k}, \eta) \varphi_a ({\bf k}, \eta ) 
   + i \int d\eta K_a (-{\bf k}, \eta) \chi_a ({\bf k}, \eta ) \right] \ ,
\end{eqnarray}
where ${\bf L}_{lin}$ is the linear part of ${\bf L}$ and $S_{int}$ denotes
the interaction part of the action. 
As the field $\varphi_a $ is linear in the action, it is easy to integarte out it as
\begin{eqnarray}       
  \exp i S_{int} [-i \frac{\delta}{ \delta J_a } , -i \frac{\delta}{\delta K_b}]
       \int  D\chi_a  ({\bf k}, \eta) e^{C[\chi_a ({\bf k}, 0) ] }
       \delta_D (\chi_a {\bf L}_{lin} + J_a )
   \exp \left[  
    i \int d\eta K_a ({\bf k}, \eta) \chi_a ({\bf k}, \eta )\right] \ ,
\end{eqnarray}
where $\chi_a {\bf L}_{lin} \equiv  \chi_b (\eta_b )
(-\overleftarrow{\partial}_{\eta_b} \delta_{ba} + \Omega_{ba} )$.
The constraint imposed by the Dirac delta function can be solved by
using the equation
\begin{eqnarray}
  g_{ab} (\eta_a , \eta_b ) 
  (-\overleftarrow{\partial}_{\eta_b} \delta_{bc} + \Omega_{bc} )
  = \delta_{ac} \delta (\eta_a - \eta_b ) 
\end{eqnarray}
as
\begin{eqnarray}
\chi_a ({\bf k}, \eta_b ) 
= -\int d\eta J_a ({\bf k}, \eta_a ) g_{ab} (\eta_a , \eta_b ) \ .
\end{eqnarray}
Now, it is straightforward to integrate  out the field $\chi_a$. 
The final result is given by
\begin{eqnarray}   
Z[J_a ,K_b ]&=& 
   \exp i S_{int} [-i \frac{\delta}{ \delta J_a } , -i \frac{\delta}{\delta K_b }]
   \exp \left[  C[-\int d\eta J_a ({\bf k}, \eta) g_{ab} (\eta , 0) ]  \right.  \nonumber\\
   && \qquad \qquad\qquad\qquad\qquad\qquad\qquad
   \left. - i \int d\eta J_a (-{\bf k}, \eta_a ) 
   g_{ab} (\eta_a , \eta_b ) K_b ({\bf k}, \eta_b ) \right] .
\end{eqnarray}
This leads to Feynman rules for calculating various quantities. 
Thus, we have shown that the graphical method is applicable to the
evolution problem of 
cosmological fluctuations with the general non-Gaussian statistics. 

It is useful to see a more concrete expression for our path integral
representation. 
In the case of the Gaussian statistics, the cumulant is  determined solely by
the power spectrum. Therefore, we have a simple expression
\begin{eqnarray}
   Z [J_a , K_b ]  = \exp \left[ -i \int d\eta e^\eta 
   \gamma_{abc}({\bf k},{\bf p},{\bf q})
    \left(-i\frac{\delta}{\delta K_a ({\bf k})} \right)
    \left(-i\frac{\delta}{ \delta J_b ({\bf p})} \right)
    \left(-i \frac{\delta}{ \delta J_c ({\bf q})}  \right)
   \right] Z_0 [J_a , K_b] ,
\label{partition}
\end{eqnarray}
where
\begin{eqnarray}
Z_0 [J_a ,K_b] &=& 
   \exp \left[  -\frac{(2\pi)^3}{2} \int d\eta_a d\eta_b J_a (-{\bf k}, \eta_a )
                P_{ab} (\eta_a , \eta_b ) J_b ({\bf k}, \eta_b ) \right.\nonumber\\
 && \left. \qquad  - i \int d\eta_a d\eta_b
    J_a (-{\bf k}, \eta_a ) g_{ab} (\eta_a , \eta_b ) K_b ({\bf k}, \eta_b ) \right] .
\end{eqnarray}
This result coincides with the expression obtained by Matarrese and Pietroni. 
In the case of the non-Gaussian statistics, we have infinite series of
the irredecible correlation functions.     
In this non-Gaussian cases, the extra contributions read
\begin{eqnarray}
Z_0 [J_a ,K_b ] &=& 
   \exp \left[ (2\pi)^3 \left( -\frac{1}{2} \int d\eta_a d\eta_b J_a (-{\bf k}, \eta_a )
                P_{ab}^L (\eta_a , \eta_b ) J_b ({\bf k}, \eta_b )  \right.\right.  \nonumber\\   
 && \qquad \left.\left.
     -\frac{i}{6} \int d\eta_a d\eta_b d\eta_c J_a (-{\bf k}, \eta_a ) 
     J_b (-{\bf k}, \eta_b ) J_c (-{\bf k}, \eta_c ) 
     B_{abc}^L \left( {\bf k}_1 , {\bf k}_2 , {\bf k}_3  \right) + \cdots   \right. \right) \nonumber\\  
 && \qquad \left.
       - i \int d\eta_a d\eta_b J_a (-{\bf k}, \eta_a ) 
       g_{ab} (\eta_a , \eta_b ) K_b ({\bf k}, \eta_b ) \right] ,
\end{eqnarray}
where we have defined the propagated initial correlation functions
\begin{eqnarray}
&&P^L_{ab} (\eta_a ,\eta_b ) \equiv g_{ac}(\eta_a,0)g_{bd}(\eta_b,0)P_{cd}, \\
&&B^L_{abc} \left({\bf k}_1 , {\bf k}_2 , {\bf k}_3, \eta_a ,\eta_b , \eta_c  \right) 
\equiv g_{ad}(\eta_a,0)g_{be}(\eta_b,0)g_{cf}(\eta_c,0) 
B_{def} \left( {\bf k}_1 , {\bf k}_2 , {\bf k}_3  \right) \ .
\end{eqnarray}
The formula (\ref{partition}) gives Feynman rules. 
In this paper, we will be concentrated on the effect of the initial 
bispectrum $B_{abc}$.

In the quantum field theory, 
the generator of the connected Green functions can be written as
\begin{eqnarray}
W [J_a ,K_b ]=-i \log Z[J_a , K_b ]     \ . \label{W}
\end{eqnarray}
The expectation values  $\phi_a$ and $\chi_a$ can be defined as
\begin{eqnarray}
\varphi_a = \frac{\delta W}{\delta J_a}, \qquad \chi_a = \frac{\delta W}{\delta K_a}.
\end{eqnarray}
Motivated from the linear expression
\begin{eqnarray}
  (2\pi)^3 P_{ab}^L = -i \frac{\delta^2 W_0 }{\delta J_a \delta J_b } \ ,
  g_{ab} = - \frac{\delta^2 W_0 }{\delta J_a \delta K_b } \ ,
  (2\pi)^3B_{abc}^L =  \frac{\delta^3 W_0 }{\delta J_a \delta J_b \delta J_c} \ ,
\end{eqnarray}
we define
\begin{eqnarray}
  (2\pi)^3 P_{ab} =-i \frac{\delta^2 W }{\delta J_a \delta J_b } \ ,
  G_{ab} = - \frac{\delta^2 W }{\delta J_a \delta K_b } \ ,
  (2\pi)^3 B_{abc} = \frac{\delta^3 W }{\delta J_a \delta J_b \delta J_c} \ .
\end{eqnarray}

Carrying out the Legendre transformation of $W$,
we can define the effective action
\begin{eqnarray}
\Gamma[\varphi_a,\chi_a]=W[J_a[\varphi_a,\chi_a],K_a[\varphi_a,\chi_a]] -\int d\eta d^3 
{\bf k} (J_a [\varphi_a,\chi_a] \varphi_a +K_b[\varphi_a,\chi_a] \chi_b ).
\label{EA}
\end{eqnarray}
where
\begin{eqnarray}
J_a[\varphi_a,\chi_a] =- \frac{\delta \Gamma}{\delta \phi_a} \ , \quad
K_a[\varphi_a,\chi_a] =- \frac{\delta \Gamma}{\delta \chi_a} \ .
\end{eqnarray}

\subsection{Renormalization Group}

% }'Ì'}"ü
\begin{figure}[tbp]
  \begin{center}
    \includegraphics[keepaspectratio=true,height=30mm]{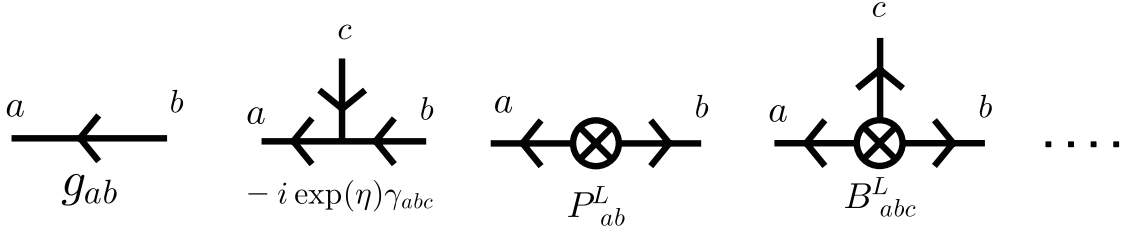}
  \end{center}
  \caption{Feymann diagrams; an arrow represents the direction of the time.}%{}"à'Ƀ^ƒCƒgƒ‹'ð‹L"ü'µ'Ä'­'¾'³'¢
  \label{fig:Renormalizedfig1.pdf}
\end{figure}

 The idea of the renormalization group introduced by Matarrese and Pietroni is as follows. 
First, we introduce a filter function with the UV cutoff $\lambda$
 in the $P^0_{ab}$ and $B^0_{abc}$.
 This defines a fictious theory where the linear perturbation theory
 works well. 
We denote various quantities in the cutoff theory with the suffix $\lambda$,
 for example like as $P_{ab,\lambda}$.
 When this cutoff scale $\lambda$ goes to infinity,
the original theory is recovered. As $\lambda$ becomes large, nonlinear
effects are incorporated gradually. This process can be expressed by
the renormalization group equation. 

The RG equation for the effective action can be deduced as follows.
Taking the derivative of the generating functional $Z_\lambda $ 
with respect to $\lambda$, we obtain
\begin{eqnarray}
\partial_\lambda Z_\lambda &=& 
\frac{(2\pi)^3}{2} 
\int d \eta_a d \eta_b \partial_\lambda P_{ab,\lambda} (p)
\delta(\eta_a) \delta(\eta_b) 
\frac{\delta^2 Z_\lambda}{\delta K_a (-{\bf p},\eta) \delta K_b ({\bf p},\eta)} 
                                                       \nonumber \\
&& \quad + \frac{(2\pi)^3}{6} \int d \eta_a d \eta_b \eta_c \partial_\lambda 
B_{abc,\lambda}({\bf p}_1 , {\bf p}_2 , {\bf p}_3 )
\delta(\eta_a) \delta(\eta_b) \delta(\eta_c ) 
\frac{\delta^3 Z_\lambda}{\delta K_a ({\bf p}_1 , \eta)
            \delta K_b ({\bf p}_2 , \eta)\delta K_c ({\bf p}_3 , \eta)}\nonumber \\
&&  \qquad\qquad\qquad\qquad\qquad\qquad\qquad\qquad\qquad\qquad
+ \ \cdots  ,
\end{eqnarray}
where we have replaced the auxiliary fields $\chi_a$ by the functional 
derivative with repect to $K_a$. The above equation can be translated to
the equation for the generating functional of the connected Green functions. 
Using the relation (\ref{W}), we obtain
\begin{eqnarray}
\partial_\lambda W_\lambda &=&\frac{(2\pi)^3}{2} 
\int d \eta_a d \eta_b \partial_\lambda P_{ab,\lambda}
\delta(\eta_a) \delta(\eta_b)
\left( \frac{\delta^2 W_\lambda}{\delta K_a (-{\bf p},\eta) \delta K_b ({\bf p},\eta)} 
+i\frac{\delta W}{\delta K_a (-{\bf p},\eta)}
    \frac{\delta W}{\delta K_b ({\bf p},\eta)} \right) \nonumber \\
&& \quad + \frac{(2\pi)^3}{6} \int d \eta_a d \eta_b d\eta_c \partial_\lambda 
  B_{abc,\lambda} ({\bf p}_1 , {\bf p}_2 , {\bf p}_3 )
\delta(\eta_a) \delta(\eta_b) \delta(\eta_c ) \nonumber \\
&&\quad \times
\left( \frac{\delta^3 W}{\delta K_a ({\bf p}_1 , \eta) 
           \delta K_b ({\bf p}_2 , \eta) \delta K_c ({\bf p}_3 , \eta)}
+ i \left( \frac{\delta W}{\delta K_a ({\bf p}_1 , \eta)}
     \frac{\delta^2 W}{\delta K_b ({\bf p}_2 , \eta) \delta K_c ({\bf p}_3 , \eta)} 
+(\mbox{cyc}.[a,b,c]) \right) \right. \nonumber \\
&& \qquad \left. - \frac{\delta W}{\delta K_a ({\bf p}_1 , \eta)}
  \frac{\delta W}{\delta K_b ({\bf p}_2 , \eta)}
  \frac{\delta W}{\delta K_c ({\bf p}_3 , \eta)} \right)
 + \ \cdots . \label{WRG}
\end{eqnarray}
Now, we can write down the RG equation for the effective action.
Using the definition (\ref{EA}), we get 
\begin{eqnarray}
&&\partial_\lambda \Gamma_\lambda [\varphi_a ,\chi_b] \nonumber \\
&&\qquad
=\partial_\lambda W_\lambda [J_{\lambda,a}[\varphi_a ,\chi_b],K_{\lambda,a}[\varphi_a ,\chi_b]]
- (\partial_\lambda J_{\lambda,a}[\varphi_a,\chi_a] \varphi_a 
+\partial_\lambda K_{\lambda,b}[\varphi_a,\chi_a] \chi_b )
\nonumber \\
&&\qquad
=\biggl\{
\frac{(2\pi)^3}{2} \int d \eta_a d \eta_b \partial_\lambda P_{ab,\lambda}
\delta(\eta_a) \delta(\eta_b)
\left( \frac{\delta^2 W_\lambda}{\delta K_a (-{\bf p},\eta) \delta K_b ({\bf p},\eta)} 
+i\frac{\delta W}{\delta K_a (-{\bf p},\eta)}
\frac{\delta W}{\delta K_b ({\bf p},\eta)} \right) \nonumber \\
&& \qquad\qquad \quad
 +\frac{(2\pi)^3}{6}  \int d \eta_a d \eta_b d \eta_c \partial_\lambda 
 B_{abc,\lambda} ({\bf p}_1 , {\bf p}_2 , {\bf p}_3 )
\delta(\eta_a) \delta(\eta_b) \delta(\eta_c ) \nonumber \\
&&\qquad\qquad \times
\left( \frac{\delta^3 W}{\delta K_a ({\bf p}_1 , \eta)
\delta K_b ({\bf p}_2 , \eta) \delta K_c ({\bf p}_3 , \eta)}
+ i \left( \frac{\delta W}{\delta K_a ({\bf p}_1 , \eta)}
   \frac{\delta^2 W}{\delta K_b ({\bf p}_2 , \eta) \delta K_c ({\bf p}_3 , \eta)} 
+(\mbox{cyc}.[a,b,c]) \right) \right. \nonumber \\
&& \quad\qquad\qquad \left.  - \frac{\delta W}{\delta K_a ({\bf p}_1 , \eta)}
\frac{\delta W}{\delta K_b ({\bf p}_2 , \eta)}
\frac{\delta W}{\delta K_c ({\bf p}_3 , \eta)} \right)
 + \ \cdots
\biggr\} \Bigg| _ {J_a=\frac{\delta\Gamma}{\delta \varphi_a},
K_a=\frac{\delta\Gamma}{\delta \chi_a}} 
\nonumber \\
&&\qquad\qquad\qquad\qquad
+ \frac{\delta W}{\delta J_{\lambda_a}}\partial_\lambda J_{\lambda,a}
+ \frac{\delta W}{\delta K_{\lambda_a}}\partial_\lambda K_{\lambda,a}
- (\partial_\lambda J_{\lambda,a}[\varphi_a,\chi_a] \varphi_a 
+\partial_\lambda K_{\lambda,b}[\varphi_a,\chi_a] \chi_b ).\label{GammaRG1}
\end{eqnarray}
The last line vanishes due to the definition of the expectation values. 
It is convenient to separate out the contribution of the tree
part from the effective action. 
Let us define the tree part
\begin{eqnarray}
&&\Gamma_{0,\lambda}\equiv
 \int d\eta_a d\eta_b  
\left\{ \frac{i(2\pi)^3}{2} \chi_a (-{\bf p},\eta) P_{ab,\lambda}(q) \delta(\eta_a)
\delta(\eta_b) \chi_b ({\bf p},\eta) + \chi_a (-{\bf p},\eta) g^{-1}_{ab}\varphi_b ({\bf p},\eta)
\right\} \nonumber\\
&&\qquad\qquad\qquad
+\frac{(2\pi)^3}{6} \int d\eta_a d\eta_b d\eta_c   
B_{abc ,\lambda} ({\bf p}_1 , {\bf p}_2 , {\bf p}_3 )
\delta(\eta_a) \delta(\eta_b) \delta(\eta_c)
\chi_a (-{\bf p}_1 ,\eta) \chi_b (-{\bf p}_2 ,\eta) \chi_c (-{\bf p}_3 ,\eta)
\end{eqnarray}
and the interaction part
\begin{eqnarray}
\Gamma_{int,\lambda} \equiv
\Gamma_\lambda -\Gamma_{0,\lambda} \ .
\end{eqnarray}
Then,  Eq.(\ref{GammaRG1}) can be written as
\begin{eqnarray}
&&\partial_\lambda \Gamma_{int,\lambda} [\varphi_a ,\chi_b] \nonumber \\
&&\qquad
=\biggl\{
\frac{(2\pi)^3}{2} \int d \eta_a d \eta_b \partial_\lambda P_{ab,\lambda}
\delta(\eta_a) \delta(\eta_b)
\frac{\delta^2 W_\lambda}{\delta K_a  (-{\bf p},\eta)\delta K_b  ({\bf p},\eta)} 
                                                 \nonumber \\
&& \qquad 
 +\frac{(2\pi)^3}{6} \int d \eta_a d \eta_b d \eta_c \partial_\lambda 
 B_{abc,\lambda}({\bf p}_1 , {\bf p}_2 , {\bf p}_3 )
\delta(\eta_a) \delta(\eta_b) \delta(\eta_c ) \nonumber \\
&&\qquad \times
\left(  \frac{\delta^3 W}{\delta K_a ({\bf p}_1 ,\eta) 
\delta K_b ({\bf p}_2 ,\eta) \delta K_c ({\bf p}_3 ,\eta)}
+ i  \frac{\delta W}{\delta K_a ({\bf p}_1 ,\eta)}
\frac{\delta^2 W}{\delta K_b ({\bf p}_2 , \eta)\delta K_c ({\bf p}_3 ,\eta)} 
+(a\leftrightarrow b \leftrightarrow c) \right) \nonumber \\
&& \qquad\qquad\qquad\qquad\qquad\qquad\qquad\qquad
\qquad\qquad\qquad\qquad
 + \ \cdots
\biggr\} \Bigg| _ {J_a=\frac{\delta\Gamma}{\delta \varphi_a}, 
K_a=\frac{\delta\Gamma}{\delta \chi_a}} \label{GammaRG}
\end{eqnarray}
This is the RG equation for general Green functions. 
In the next section, we will apply our formalism to
the calculation of the nonlinear propagator.

%===============================================================%
%************************ SECTION IV **************************%
%===============================================================%
\section{Nonlinear propagator in the presence of non-Gaussianity}

Nonlinear interaction causes a deviation from the linear propagator
$g_{ab}$. This effect is interpreted as a propagator renormalization
by Crocce and Scoccimarro~\cite{Crocce:2005xz}. Physicaly, it is interpreted as a
measure of the memory of initial conditions.
The nonlinear propagator is defined by 
\begin{eqnarray}
  G_{ab} ( k , \eta) 
            = \frac{<\varphi_a ({\bf k} , \eta ) \varphi_b  (-{\bf k}, 0) >}
                 {<\varphi_a ({\bf k} , 0 ) \varphi_b  (-{\bf k}, 0) >} \ .
\end{eqnarray}
If the linear approximation is good, the above expression should give
the linear propagator $g_{ab} (\eta)$. 

We shall calculate the nonlinear propagator by employing
the renormalization group equation. 
Doubly differentiating Eq.(\ref{WRG}) with respect to 
$K_a$ and $J_a$, 
we can get the RG equation for the propagator as
\begin{eqnarray}
&&\partial_\lambda \frac{\delta^2 W_\lambda}
{\delta J_a(-{\bf k},\eta_a)\delta K_b ({\bf k'},\eta_a)}
=-\delta({\bf k} - {\bf k'}) \partial_\lambda 
G_{ab,\lambda}(k;\eta_a \eta_b) \nonumber \\
&&\qquad
=\biggl\{
\frac{(2\pi)^3}{2} \int d \eta_c d \eta_d \partial_\lambda P_{cd,\lambda}
\delta(\eta_c) \delta(\eta_d)
\frac{\delta^4 W_\lambda}{\delta J_a (-{\bf k},\eta_a)
\delta K_b ({\bf k}' ,\eta_a) 
\delta K_c (-{\bf p},\eta_a) \delta K_d ({\bf p},\eta_a)} \nonumber \\
&& \qquad\qquad \quad
 +\frac{(2\pi)^3}{6} \int d \eta_c d \eta_d d \eta_e \partial_\lambda 
 B_{cde,\lambda}({\bf p}_1 , {\bf p}_2 , {\bf p}_3 )
\delta(\eta_c) \delta(\eta_d) \delta(\eta_e )  \nonumber \\
&& \qquad\qquad \quad \times
 \frac{\delta^5 W_\lambda }{\delta J_a (-{\bf k},\eta_a) \delta K_b ({\bf k}', \eta_a)
 \delta K_c (-{\bf p}_1, \eta_a) \delta K_d (-{\bf p}_2, \eta_a) 
 \delta K_e (-{\bf p}_3 ,\eta_a)}          \nonumber \\
&& \qquad\qquad\qquad\qquad\qquad\qquad\qquad\qquad
\qquad\qquad\qquad\qquad
 + \ \cdots
\biggr\} \Bigg| _ {J_a=0, K_a=0} .
\end{eqnarray}

In the above, we have used the fact~\footnote{
This can be proved as follows. First of all, we need to keep
it in our mind that the propagator is causal.  
Suppose that we take derivatives of $W$ with respect to only $K_a$ (not $J_a$), 
then propagators of all external lines go to the future direction.
Now, let us consider the propagator from the most future external line. 
This propagator must connect to a vertex.
However, one of the edges of the vertex still directs the future.
This means that the external line is needed in the future,
which contradicts the initial assumption.} 

\begin{eqnarray}
  \frac{\delta^n W[J_a , K_b ]}{\delta K_{a_1} (k_1 , \eta_1 ) \cdots 
  \delta K_{a_n} (k_n , \eta_n)}
  \big|_{J_a =0, K_a=0} = 0\ ,   \quad {\rm for \  any}\  n .
\end{eqnarray}

Now, we shall rewrite connected correlation functions by the irreducible
correlation functions
\begin{eqnarray}
  \delta_D ( k_1 + \cdots + k_n)
   \Gamma^{(n)}_{\varphi_{a_1} \chi_{a_2} \cdots \chi_{a_n}} 
  = \frac{\delta^n \Gamma[\varphi_a , \chi_b ]}{\delta \chi_{a_1} (k_1 , \eta_1 ) 
  \cdots \delta \varphi_{a_n} ( k_n , \eta_n)}
  \big|_{\varphi_a =0, \chi_b =0}  \ .
\end{eqnarray}
%
% }'Ì'}"ü
\begin{figure}[tbp]
  \begin{center}
    \includegraphics[keepaspectratio=true,height=60mm]{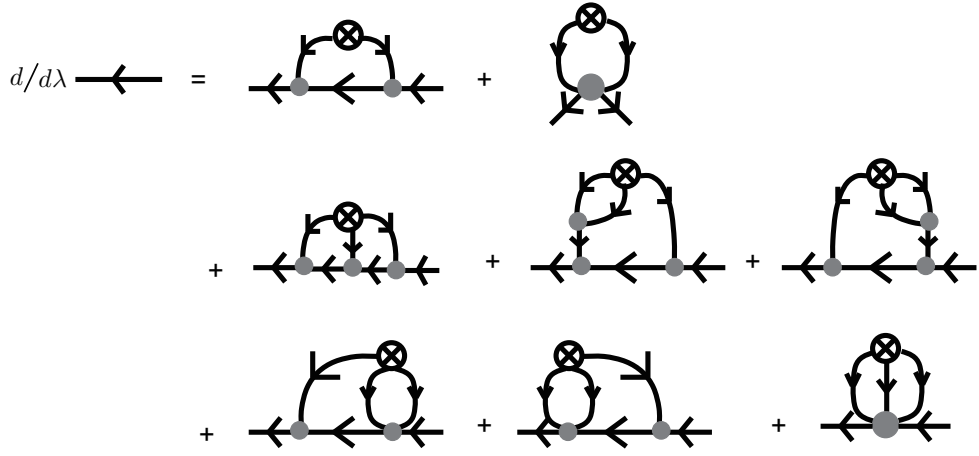}
  \end{center}
  \caption{RG equation diagram for the propagator.}%{}"à'Ƀ^ƒCƒgƒ‹'ð‹L"ü'µ'Ä'­'¾'³'¢
  \label{fig:Renormalized-CPTfig2.eps}
\end{figure}
%KI
For example, the four-point function can be written as
\begin{eqnarray}
&&\frac{\delta^4 W_\lambda}{\delta J_a \delta K_b\delta K_c \delta K_d}\nonumber\\
&&
=\int ds_1 \cdots ds_4 G_{ae,\lambda}(k_1;\eta_a,s_1)
G_{fb,\lambda}(k_2;s_2,\eta_b)G_{gc,\lambda}(k_3;s_3,\eta_c)
G_{hd,\lambda}(k_4;s_4,\eta_d)\nonumber\\
&&\qquad\times\bigl\{
-\int ds_5 ds_6 G_{ij,\lambda}(k_1-k_2;s_5,s_6)\nonumber \\
&&\qquad\qquad\qquad\times
\Gamma^{(3)}_{\chi_e\varphi_f \varphi_i}(k_1,s_1;k_2,s_2;k_1-k_2,s_5)
\Gamma^{(3)}_{\chi_j\varphi_j \varphi_h}(k_1-k_2,s_6;k_3,s_3;k_4,s_4)  \nonumber\\
&&\qquad\qquad
+\Gamma^{(4)}_{\chi_e\varphi_f \varphi_g\varphi_h}(k_1,s_1;k_2,s_2;k_3,s_3;k_4,s_4) \bigr\}
 \delta(k_1-k_2-k_3-k_4).
\end{eqnarray}
In Fig.\ref{fig:Renormalized-CPTfig2.eps}, 
we have drawn the Feynman diagrams for the RG equation.
The four-point function in the RG equation gives the first and 
second diagrams in Fig.\ref{fig:Renormalized-CPTfig2.eps}. 
Other diagrams in Fig.\ref{fig:Renormalized-CPTfig2.eps} come from the five-point function
\begin{eqnarray}
&&\frac{\delta^5 W_\lambda}{\delta J_a \delta K_b\delta K_c \delta K_d \delta K_e}\nonumber\\
&&
=\int ds_1 \cdots ds_4 G_{af,\lambda}(k_1;\eta_a,s_1)
G_{gb,\lambda}(k_2;s_2,\eta_b)G_{hc,\lambda}(k_3;s_3,\eta_c)
G_{id,\lambda}(k_4;s_4,\eta_d)G_{je,\lambda}(k_5;s_5,\eta_e)\nonumber\\
&&\qquad\times\bigl\{
\int ds_6 ds_7ds_8ds_9 G_{kl,\lambda}(k_1-k_2;s_6,s_7)
G_{mn,\lambda}(k_5+k_4;s_8,s_9)
\Gamma^{(3)}_{\chi_f\varphi_g \varphi_k}(k_1,s_1;k_2,s_2;k_1-k_2,s_6)\nonumber \\
&&\qquad\qquad\qquad\times
\Gamma^{(3)}_{\chi_l\varphi_h \varphi_m}(k_1-k_2,s_7;k_3,s_3;k_5+k_4,s_8)
\Gamma^{(3)}_{\chi_n\varphi_i \varphi_j}(k_5+k_4,s_8;k_4,s_4;k_5,s_5)  \nonumber\\
&&\qquad\qquad\qquad\qquad\qquad\qquad\qquad\qquad\qquad\qquad\qquad\qquad\qquad\qquad
+ \cdots  \bigr\}
 \delta(k_1-k_2-k_3-k_4-k_5) \ , 
\end{eqnarray}
where we have explicitly displayed only the relevant one.

We can now calculate the RG equation. 
On the nonlinear scale, modes enter into the  damping phase.
Therefore, the nonlinear effect allows us to
consider only the low wavenumber mode for the internal lines.  
Hence, we can take the large $k$ limit,
where $k$ is the wavenumber of the renomalized propagator. 
In this paper, we approximate a solution by taking into account 
only the running of the 2-point function. We
 also ignore the propagation of the decaying mode.
Thus, the coupling constant $\gamma_{abc}$ always appears 
in the following form:
\begin{eqnarray}
\gamma ({\bf k},{\bf p},{\bf q}) &\equiv&
\hat u_a  \gamma_{abc}({\bf k},{\bf p},{\bf q}) 
u_b u_c \nonumber\\
&=&
\frac{1}{10p^2q^2}(6p^2q^2 +5(p^2+q^2){\bf p}\cdot{\bf q}
+4({\bf p}\cdot{\bf q})^2)  \ .
\label{gamma}
\end{eqnarray}
In the limit $p \gg q$, we obtain
\begin{eqnarray}
\gamma ({\bf k},{\bf p},{\bf q})
  \xrightarrow[p \gg q]{}
\frac{{\bf p}\cdot{\bf q}}{2q^2} \ .
\label{gamma1}
\end{eqnarray}
In our approximation, 
the diagrams  we need to consider are the first, third, fourth 
and fifth ones in Fig.\ref{fig:Renormalized-CPTfig2.eps}.
It is known that the fourth and fifth ones are higher order 
contributions than the third one~\cite{Crocce:2005xz}. 
The third one directly interacts with the initial conditions,
so that all informations of the intial conditions transmit.
From Eq.(\ref{gamma}), we see the direct interaction linearly 
grows with the wavenumber ${\bf p}$.
On the contrary, there are interactions outside the main path
in the fourth and fifth diagrams.
There, we can not take the limit $k\gg q$ and, 
hence, we cannot expect the linear growth.
Thus, we can ignore them.
%However, when we finally calculate the power spectrum, 
%we can take into account their contributions 
%through the running effect of the 1PI two-point function.
As an additional approximation, we replace the propagator in the right hand side of 
the RG equation in Fig.\ref{fig:Renormalized-CPTfig2.eps}
with its linear expression.

Using the above aproximations,
we can get the RG eqation for the propagator,
\begin{eqnarray}
&&\partial_\lambda G_{ab,\lambda}(k;\eta_a,\eta_b) 
\nonumber\\
&&
=g_{ab}(\eta_a,\eta_b)
\biggl\{ 
2^2 (2\pi)^3 
\int^{\eta_a}_{\eta_b} \!\!\!\! ds_2
\int^{s_2}_{\eta_b} \!\!\!\! ds_1 \exp(s_1+s_2)
\nonumber\\
&&\qquad\qquad\qquad\qquad\times 
\int d^3{\bf q} 
 \partial_\lambda P_{cd,\lambda} (q) \hat u_c \hat u_d
\gamma({\bf k},{\bf q},{\bf k}-{\bf q})
\gamma({\bf k}-{\bf q},-{\bf q},{\bf k})
\nonumber\\
&&\qquad\qquad\qquad\qquad
-2^3(2\pi)^3 \int^{\eta_a}_{\eta_b} \!\!\!\! ds_3
\int^{s_3}_{\eta_b} \!\!\!\! ds_2
\int^{s_2}_{\eta_b} \!\!\!\! ds_1 \exp(s_1+s_2+s_3)
\nonumber\\
&&\qquad\qquad\qquad\qquad\qquad\times
 \int d^3{\bf q_1}d^3{\bf q_2}d^3{\bf q_3}
 \partial_\lambda B_{cde,\lambda} ({\bf q_1},{\bf q_2},{\bf q_3})
  \hat u_c \hat u_d  \hat u_e  
\nonumber\\
&&\qquad\qquad\qquad\qquad\qquad\quad\times
\gamma({\bf k},{\bf q_1},{\bf k}-{\bf q_1})
\gamma({\bf k}-{\bf q_1},{\bf k}+{\bf q_3},{\bf q_2})
\gamma({\bf k}+{\bf q_3},{\bf q_3},{\bf k})
\biggr\}
\nonumber\\
&&
=g_{ab}(\eta_a,\eta_b)
\biggl\{ 
2\left(
\exp(\eta_a)- \exp(\eta_b)
\right) ^2 \partial_\lambda F_{P,\lambda}[k] 
-8\left(
\exp(\eta_a)- \exp(\eta_b)
\right) ^3 \partial_\lambda F_{B,\lambda}[k]
\biggr\} \ , \label{diff}
\end{eqnarray}
where we have defined 
\begin{eqnarray}
F_{P,\lambda}[k]& \equiv& 
(2\pi)^3\int d^3{\bf q} 
 P_{cd,\lambda} (q) \hat u_c  \hat u_d
\gamma({\bf k},{\bf q},{\bf k}-{\bf q})
\gamma({\bf k}-{\bf q},-{\bf q},{\bf k})\nonumber\\
&=&(2\pi)^3 \frac{\pi}{50} \int  dq
 P_{cd,\lambda} (q) \hat u_c  \hat u_d
\biggl( 3\frac{k^4}{q^2} -\frac{121}{6}k^2 +9q^2 -\frac{9}{2} \frac{q^4}{k^2}
\nonumber\\
&&\qquad\qquad\qquad
+\left(-\frac{3}{2} \frac{k^5}{q^3} +\frac{9}{4}\frac{k^3}{q} +\frac{9}{4}k q 
-\frac{21}{4} \frac{q^3}{k}
+\frac{9}{4}\frac{q^5}{k^3} \right)\log\left|\frac{k+q}{k-q}\right|\biggr)
\end{eqnarray}
and
\begin{eqnarray}
F_{B,\lambda}[k]& \equiv&
 (2\pi)^3\int d^3{\bf q_1}d^3{\bf q_2}d^3{\bf q_3}
  B_{cde,\lambda} ({\bf q_1},{\bf q_2},{\bf q_3})
   \hat u_c  \hat u_d  \hat u_e  
\nonumber\\
&&\qquad\qquad\quad\times
\gamma({\bf k},{\bf q_1},{\bf k}-{\bf q_1})
\gamma({\bf k}-{\bf q_1},{\bf k}+{\bf q_3},{\bf q_2})
\gamma({\bf k}+{\bf q_3},{\bf q_3},{\bf k}).
\end{eqnarray}
Here, we promote the linear propagator to the full nonlinear one
following Mataresse and Pietroni~\cite{Matarrese:2007wc}. 
Solving the differential equation~(\ref{diff}) 
with the initial condition $G_{ab,\lambda=0}=g_{ab}$
and taking the limit $\lambda$ goes to infinity, 
we obtain
\begin{eqnarray}
&&G_{ab}(k;\eta_a,\eta_b)
= g_{ab}(k;\eta_a,\eta_b)
\nonumber\\
&&\qquad\times
\exp\Bigl\{ 
2\left(
\exp(\eta_a)- \exp(\eta_b)
\right) ^2 F_{P}[k] 
-8\left(
\exp(\eta_a)- \exp(\eta_b)
\right) ^3 F_{B}[k]
\Bigr\}                   \label{nonpro}
\end{eqnarray}

As a concrete example, we consider the model
\begin{eqnarray}
\varphi_0 ({\bf x}) = \varphi_g ({\bf x}) 
+ \xi_{NL} (\varphi_g^2 ({\bf x}) - \langle \varphi_g^2 \rangle ) ,
\end{eqnarray}
where $\varphi_g ({\bf x})$ is a Gaussian field.
It should be noted that this $\xi_{NL}$
is different from $f_{NL}$, because we are considering
$\varphi_a$ instead of the curvature perturbation $\Phi$. 
Then the initial bispectrum $B_{abc}$ is written as
\begin{eqnarray}
B_{abc}({\bf q_a},{\bf q_b},{\bf q_c}) 
= 2 \xi_{NL}
 \delta({\bf q_a}+{\bf q_b}+{\bf q_c})
 (P(q_a)P(q_b) +P(q_b)P(q_c) +P(q_c)P(q_a)) ,
\end{eqnarray}
where $P(q) = P_{cd} (q) \hat u_c \hat u_d$.
In the $k \gg p$ limit, we can deduce 
\begin{eqnarray}
F_{P}[k]\xrightarrow[k \gg q]{} 
   - (2\pi)^3\frac{\pi k^2}{3} \int d q 
P_{cd} (q) \ , 
\end{eqnarray}
and
\begin{eqnarray}
&&F_{B}[k] 
\xrightarrow[k \gg q]{}
   (2\pi)^3 \frac{7}{80} \int d^3 q_1 d^3 q_2 d^3 q_3 \nonumber\\
&& \qquad\qquad\qquad\qquad\times    
    \left(
    \frac{({\bf k}\cdot{\bf q_1})({\bf k}\cdot{\bf q_2})}{q_1^2 q_2^2}
   +\frac{({\bf k}\cdot{\bf q_2})({\bf k}\cdot{\bf q_3})}{q_2^2 q_3^2}
   +\frac{({\bf k}\cdot{\bf q_3})({\bf k}\cdot{\bf q_1})}{q_3^2 q_1^2}
                                               \right) 
    B ({\bf q}_1 , {\bf q}_2 , {\bf q}_3 ) \ ,
%&&F_{B}[k] 
%\xrightarrow[k \gg q]{}
%    \frac{4 k^2}{45} \int d^3 q_1 d^3 q_2 d^3 q_3 \frac{q_1 + q_2 + q_3}{q_1 q_2 q_3} 
%    B ({\bf q}_1 , {\bf q}_2 , {\bf q}_3 ) \ ,
\end{eqnarray}
where we have defined $B(q) = B_{cde} (q) \hat u_c \hat u_d \hat u_e$.

From the shape of the CDM spectrum, we can estimate the scale 
where the running of the propagator is important.
The nonlinear effect works 
when the argument of exponential in Eq.(\ref{nonpro}) becomes ${\cal O}(1)$.
The threshold for the power spectrum ($F_P a_0^2$)
becomes $k\sim 0.1(1+z)\mbox{Mpc}^{-1}$.
On the other hand,
the threshold for the bispectrum ($F_B a_0^2$)
is given by $k\sim0.03 \xi_{NL}^{-{1\over2}} (1+z)^{3/2} \mbox{Mpc}^{-1}$.
This means that, if the statistics is positively skewed and $\xi_{NL} > O(1)$,
the effect of the initial bispectrum is a dominant one at present 
in the non-linear propagator.
On the other hand, the negatively skewed non-Gaussianity
delays the time each comoving mode enters the nonlinear regime. 
The above arguments imply that the non-Gaussianiy significantly 
affects the running of the propagator.

The propagator can be regarded as the measure of the memory of initial conditions.
Only in the case the memory is left in the data, 
we can detect the primordial non-Gaussianity. According to the behavior
of the nonlinear propagator, the modes in the range $k> 0.1 (1+z) \mbox{Mpc}^{-1}$
or $k>0.03 \xi_{NL}^{-{1\over2}} (1+z)^{3/2} \mbox{Mpc}^{-1}$
seems to have already lost the memory to the initial conditions.
This means that, at the time $z \sim 1$, we can not observe the primordial non-Gaussianity 
around the scale $k=1 \mbox{Mpc}^{-1}$.
In order to investigate the primordial non-Gausianity at the comoving scale $1\mbox{Mpc}$, 
we have to look at the Universe at $z >10$.
One possibility  is the observation of $21\mbox{cm}$ line. 
As to this, many observational projects, 
such as the Low Frequency Array (LOFAR), 
the Square Kilometer Array (SKA),
the Mileura Widefield Array (MWA) and 
the $21\mbox{cm}$ array (21CMA), 
are ongoing or being planed~\cite{Furlanetto:2006jb}.

%\subsection{RG equation of 1PI two-point function}
%
%As in the case of the propagator,
%we can get the RG equation of the 1PI two-point function,
%which is
%Eq.(\ref{GammaRG}) differentiated doubly with respect to $\chi_a$
%and which is written as
%%
%\begin{eqnarray}
%&&\partial_\lambda \Phi_{ab_\lambda}
%= \frac{\delta^2 \Gamma}{\delta \chi_a \delta \psi_c}
%\frac{\delta^2 \Gamma}{\delta \chi_b \delta \psi_d}
%\frac{\delta}{\delta {\bf J_c}}
%\frac{\delta}{\delta {\bf J_d}}
%\nonumber\\
%&&\qquad\qquad
%\times\biggl\{
%\frac{1}{2} \int d \eta_e d \eta_f \partial_\lambda P_{\lambda,ef}
%\delta(\eta_e) \delta(\eta_f)
%\frac{\delta^2 W_\lambda}{\delta K_e \delta K_f} 
% + \ \cdots
%\biggr\} \Bigg| _ {J_a=\frac{\delta\Gamma}{\delta \phi_a}, 
%K_a=\frac{\delta\Gamma}{\delta \chi_a}},
%\end{eqnarray}
%%
%where
%$\psi_a \equiv (\phi_a, \chi_a)$
%and
%${\bf J_a } \equiv (J_a,K_a)$.
%Its diagram is Fig~(\ref{fig:Renormalized-CPT1fig3.eps}).
%% }'Ì'}"ü
%\begin{figure}[tbp]
%  \begin{center}
%    \includegraphics[keepaspectratio=true,height=15mm]{Renormalized-CPT1fig3.eps}
%  \end{center}
%  \caption{}%{}"à'Ƀ^ƒCƒgƒ‹'ð‹L"ü'µ'Ä'­'¾'³'¢
%  \label{fig:Renormalized-CPT1fig3.eps}
%\end{figure}
%
%

%===============================================================%
%************************ SECTION V **************************%
%===============================================================%
\section{Conclusion}

We have studied the Newtonian cosmological perturbation theory
 from the field theoretical point of view.
 We have extended the renomalization group method proposed by Mataresse and Pietroni
 to the case of a primordial non-Gaussian density fluctuations. 
 In particular, we have obtained the generating functional for the cosmic
 evolution of fluctuations with non-Gaussian statistics. 
 As an application, we have calculated the nonlinear propagator
 and examined how the non-Gaussianity affects the memory of cosmic fields
 to their initial conditions. It turned out that the initial non-Gaussianity
 affects on the running of the propagator. For the positively skewed case,
 the nonlinearity starts at ealier stage. In the opposite case,
 the nonlinearity is postpond compared with the Gaussian case.

Assuming the ${\cal O}(1)$ positively skewed non-Gaussianity,
we can conclude that
the nonlinear propagator damps in the range $k> 0.1 (1+z) \mbox{Mpc}^{-1}$
or $k>0.03 \xi_{NL}^{-{1\over2}} (1+z)^{3/2} \mbox{Mpc}^{-1}$. 
Hence, if we want to investigate the initial non-gaussianity at 
the scale $k>1\mbox{Mpc}^{-1}$, the observation at $z>10$ is required.
One interesting possibility is to observe fluctuations of
$21\mbox{cm}$ absorption line in the cosmic maicrowave background radiation.
In the negatively skewed case, the effect is opposite.
Therefore, the time entering the nonlinear scale is postponed for 
the mode with a fixed comoving wavenumber. 

As an application of our formalism, we can estimate the effect of the 
primordial non-Gaussianity on the BAO. More intriguingly, we can calculate
the bispectrum of cosmic fields which provide a more clear test of non-Gaussianity.  
It is also intriguing to calculate BAO feature in the bispectrum. 
Again, the bispectrum of 21 cm line fluctuations is an interesting target.

 Recent observational progress allows us to know the large scale structure 
 at high redshift. It should be emphasized that the field theoretical
 approach gives a simple way to calculate correlation functions
with different times. Therefore, the field theoretical method in cosmology
deserves further investigations.

\begin{acknowledgements}
This work was supported in part 
  by a Grant-in-Aid for the 21st Century COE ``Center for
  Diversity and Universality in Physics". 
J.S. is supported by 
the Japan-U.K. Research Cooperative Program, the Japan-France Research
Cooperative Program and Grant-in-Aid for  Scientific
Research Fund of the Ministry of Education, Science and Culture of Japan 
  No.18540262 and No.17340075.  
\end{acknowledgements}

\end{document}